\documentclass[final,5p,times,twocolumn]{elsarticle}

\usepackage{tikz}
\usetikzlibrary{shapes.geometric, arrows.meta, positioning}

\tikzstyle{module} = [rectangle, rounded corners, minimum width=3cm, minimum height=1cm,text centered, draw=black, fill=blue!10]
\tikzstyle{arrow} = [thick,->,>=Stealth]
\usepackage{hyperref}
\hypersetup{colorlinks=true, citecolor=blue, filecolor=blue, linkcolor=blue, urlcolor=blue}
\usepackage[utf8]{inputenc}
 
\usepackage{listings}
\usepackage{xcolor}
 
\definecolor{codegreen}{rgb}{0,0.6,0}
\definecolor{codegray}{rgb}{0.5,0.5,0.5}
\definecolor{codepurple}{rgb}{0.58,0,0.82}
\definecolor{backcolour}{rgb}{0.95,0.95,0.92}
 
\lstdefinestyle{mystyle}{
    backgroundcolor=\color{backcolour},   
    commentstyle=\color{codegreen},
    keywordstyle=\color{magenta},
    numberstyle=\tiny\color{codegray},
    stringstyle=\color{codepurple},
    basicstyle=\ttfamily\footnotesize,
    breakatwhitespace=false,         
    breaklines=true,                 
    captionpos=b,                    
    keepspaces=true,                 
    numbersep=5pt,                  
    showspaces=false,                
    showstringspaces=false,
    showtabs=false,                  
    tabsize=2
}
 
\usepackage{amssymb}
\usepackage{amsmath}
\newcounter{bla}

\journal{Computer Physics Communications}

\begin{document}

\begin{frontmatter}

\title{GPR\_calculator: An On-the-Fly Surrogate Model to Accelerate Massive Nudged Elastic Band Calculations}

\author[a]{Isaac Onyango}
\author[b]{Byungkyun Kang}
\author[a]{Qiang Zhu \corref{author}}

\cortext[author] {Corresponding author.\\\textit{E-mail address:} qzhu8@charlotte.edu}
\address[a]{Department of Mechanical Engineering and Engineering Science, University of North Carolina at Charlotte, Charlotte, NC 28223, USA}

\address[b]{College of Arts and Sciences, University of Delaware, Newark, DE 19716, USA}

\begin{abstract}
We present \texttt{GPR\_calculator}, a package based on Python and C++ programming languages to build an on-the-fly surrogate model using Gaussian Process Regression (GPR) to approximate computationally expensive electronic structure calculations. The key idea is to dynamically train a GPR model during the simulation that can accurately predict energies and forces with uncertainty quantification. When the uncertainty is high, the costly electronic structure calculation is performed to obtain the ground truth data, which is then used to update the GPR model. To illustrate the effectiveness of \texttt{GPR\_calculator}, we demonstrate its application in Nudged Elastic Band (NEB) simulations of surface diffusion and reactions, achieving 3-10 times acceleration compared to pure ab initio calculations. The source code is available at \url{https://github.com/MaterSim/GPR_calculator}. 

\end{abstract}


\end{frontmatter}

\noindent
{\bf PROGRAM SUMMARY}\\
\begin{small}
\noindent
{\em Program Title:} GPR\_calculator \\
{\em Licensing provisions:} MIT \cite{1}\\
{\em Programming language:} Python 3 \& C++\\
{\em Nature of problem:} Many atomistic simulations—such as geometry optimization, barrier calculations, molecular dynamics, and equation-of-state simulations—require sampling a large number of atomic configurations in a compact phase space. While Density Functional Theory (DFT) provides good accuracy and relatively scalable performance for systems with fewer than hundreds of atoms, it can become prohibitively expensive for massive simulations. This is particularly evident in energy barrier calculations for surface diffusion or reaction studies, where hundreds or thousands of energy and force evaluations are needed. \\

{\em Solution method:} The \texttt{GPR\_calculator} is an On-the-Fly Atomistic Calculator based on Gaussian Process Regression (GPR), designed as an add-on module that can be used with the popular Atomic Simulation Environment (ASE). It is essentially a hybrid approach that consists of: (i) a base calculator to provide ground truth reference energy and forces for the given input structure, and (ii) a surrogate model serving as the less expensive approximation trained on-the-fly. When the uncertainty of the GPR prediction exceeds a user-defined threshold, the base calculator is invoked to obtain accurate results and update the GPR model. This adaptive approach ensures accuracy while significantly reducing computational cost. \\
\\

\end{small}

\section{Introduction}
\label{intro}

The role of computational modeling approaches has become increasingly important in the design of catalysts. Developing reliable computational models requires a fundamental understanding of catalytic surface reactions, particularly the relevant elementary steps and their kinetics (including the activation barriers and reaction rates). A common approach to study surface reactions is to identify the minimum energy pathway (MEP), where the energy maximum (saddle point) along this path gives the activation barrier. This barrier is then used to calculate kinetic rates within the harmonic transition state theory (HTST) framework \cite{QHTST}.

In solid state chemistry, the nudged elastic band (NEB) method \cite{NEB-2000, CINEB-2000, Herbo-NEB-2017, JONSSON-1998} is arguably the most popular approach used to search for MEPs. Generally speaking, the NEB method is a refinement of the chain of states method in which the path between two local minima is initialized as a discrete chain of configurations, commonly referred to as \textit{images}, generated by interpolation between the two local minima. The path is then optimized by iteratively minimizing an objective function defined as the sum of the energies of the images plus a Hookean energy term as a penalty that keeps the images distributed along the path through. However, optimization of this objective function tends to lead to \textit{corner cutting} where the spring force perpendicular to the path pulls images off the MEP leading to over estimation of the activation barrier if the spring constant is too large. If the spring constant is too small, the net spring force acts to balance the true force (negative gradient of the energies of the images) parallel to the path making the images slide toward lower-energy regions (\textit{sliding down}) and avoid the activation barrier. NEB alleviates the \textit{corner cutting} and \textit{sliding down} issues by projecting out the spring forces that are perpendicular to the path and the true force components parallel to the path during optimization.

In a typical NEB simulation, 5-10 images are used, with their positions iteratively adjusted during the optimization. Converging to the desired MEP often requires hundreds of optimization steps, translating to hundreds of energy and force evaluations per image using accurate electronic structure methods, most commonly based on density functional theory (DFT). The majority of computational effort is dedicated to these DFT calculations. Each electronic structure evaluation typically takes tens or even hundreds of CPU minutes, making NEB calculations very demanding. Consequently, a single NEB simulation may take several hours to days to complete. Furthermore, surface chemical reactions often involve multiple elementary steps, each requiring its own NEB calculation, further increasing computational costs. However, the accuracy of the energy landscape, where most of this computational effort is spent, is not the primary concern; rather, the key objective is to identify the saddle point. Thus, leveraging an approximate energy landscape that still enables saddle point identification could significantly reduce computational costs.

Over the past several years, various efforts have been made to accelerate NEB simulations by using the surrogate machine learning force field (MLFF) models as substitutes for expensive DFT calculations. Peterson et al. \cite{NEB-ML-2016} developed a neural network (NN) approach using the Behler-Parrinello descriptors \cite{BP-PRL-2007} to represent the local atomic environment for the images. While this leads to significantly faster simulations, it requires an iterative data collection procedure to train the model. After the model is built, it lacks an in-time validation mechanism and one cannot infer the uncertainty of the simulation results.

Jonsson et al. \cite{Jonsson-2017, Jonsson-2019} proposed a Gaussian Process Regression (GPR) model, where the covariance matrix was constructed using a radial basis function (RBF) kernel based on relative Cartesian distances and inverse interatomic distances. This approach leverages the uncertainty estimates from GPR, selectively adding only the most uncertain images to the training dataset, thereby improving the efficiency of data collection. However, the model refinement closely follows that of Peterson et al. \cite{NEB-ML-2016}, with the key distinction that only the image with the highest uncertainty estimate from the GPR is added to the training dataset after validating the final images in the GPR-optimized MEP against \textit{ab initio} calculations. 

Following Jonsson \cite{Jonsson-2017, Jonsson-2019}, several other GPR-based approaches have been introduced. In an effort to reduce the amount of training data, Torres et al. \cite{NEB-ML-2019} initiated the GPR training with an image located one-third of the way between the initial and final states in the initial MEP guess. The GPR is then used to optimize the MEP, and an image with either the highest uncertainty or the highest energy along the optimized path is selected to update the GPR model. Teng et al. \cite{Prior-Mean-GPR-2022, Prior-Mean-GPR-2024} introduced prior mean functions — either force-field-based or classical mechanical descriptions — into the GPR, in contrast to the widely used zero-mean GPR models. These prior means provide an initial estimate of the energy landscape, improving the performance of the GPR model. Furthermore, the prior means are not fixed and can be adjusted dynamically throughout the calculation. More recently, Schaaf et al. \cite{Schaaf-2023} developed a protocol aimed at creating a general MLFF capable of accurately predicting a broad range of surface phenomena, including molecular adsorption, surface rearrangement, and both single-step and multi-step reactions. Before beginning NEB or geometry optimizations, the MLFF is trained using molecular dynamics. 

Broadly speaking, all these approaches can be categorized as \textit{one-shot} MLFF models, where the MLFF is trained using different strategies, employed to optimize the MEP and subsequently updated by validating the final images against their \textit{ab initio} calculations. While such \textit{one-shot MLFFs} (either based on NN or GPR models) are becoming increasingly popular, they may not be suitable for exploratory simulation tasks that demand very high accuracy. 

To address this challenge, we propose an on-the-fly approach where MLFF models are dynamically updated during NEB calculation thus integrating both the GPR model and DFT (ab initio) calculations during MEP optimization. This approach is inspired by Jinnouchi et al. \cite{Jinnouchi-2019} who implemented the on-the-fly strategy to speed up ab initio molecular dynamics (MD) simulations for the study of phase transition, electrochemical reaction and other complex phenomena. Additionally, we integrate our approach with the Atomic Simulation Environment (ASE) \cite{ASE-2017} to ensure compatibility with a wide variety of electronic structure calculation methods besides DFT. In the following sections, we will provide a detailed description of the GPR model and its implementation in the \texttt{GPR\_calculator} package. We will also demonstrate its application in NEB simulations of surface diffusion and reactions, showcasing its ability to achieve 3-10 times acceleration compared to pure ab initio calculations.

\section{Computational Methodology}
\subsection{Gaussian Process Regression}
Suppose we have a set of sample data \{$\boldsymbol{x}, Y$\}, where $\boldsymbol{x}$ represents the input vector and $Y$ is the corresponding learning target. The GPR model assumes that the target values are drawn from a Gaussian process characterized by a kernel function $k(\boldsymbol{x}, \boldsymbol{x'})$ \cite{bishop2006pattern}. This kernel function defines the relationships between input vectors. Among various kernel functions, the RBF kernel is widely used and it is defined as:

\begin{equation}
    k(\boldsymbol{x}_m, \boldsymbol{x}_n) = \theta^2 \exp[-(\boldsymbol{x}_m - \boldsymbol{x}_n)^2/2l^2]
\end{equation}

Further, the covariance matrix $\boldsymbol{C}$ is defined as:
\begin{equation}
     \boldsymbol{C}_{mn}=\boldsymbol{C}(x_m, x_n) = k(x_m, x_n) + \beta \boldsymbol{\delta}_{mn},
\end{equation}

where $\beta$ is the noise variance, and $\boldsymbol{\delta}_{mn}$ is the Kronecker delta function. For $N$ samples, $\boldsymbol{C}$ is a square matrix of $N \times N$. Each sample value can be considered as the linear combination of these covariances.
\begin{equation}
    \boldsymbol{Y_m} = \sum_{i=1}^N \boldsymbol{G}_{\alpha i} \boldsymbol{C}(\boldsymbol{x}_m, \boldsymbol{x}_i)
\end{equation}

Hence, one only needs to determine $\boldsymbol{G}_{\alpha}$ from the previous training data. However, a computational bottleneck in implementing GP regression is the $O(N^3)$ computational complexity associated with inverting or factorizing the $\boldsymbol{C}$. When $\boldsymbol{C}_{mn}$ is smooth, the covariance matrix can be expected to be rank-deficient, i.e., its eigenvalues are likely to decay rather rapidly. A reasonable rank-$M$ approximation to $\boldsymbol{C}_N$ can be obtained by switching off its last $N-M$ eigenvalues to zero. Therefore it is possible to reduce the computational time to in $O(NM^2)$ time, by Cholesky factorization of $\boldsymbol{C}$ to its lower triangular matrix $\boldsymbol{L}$,
\begin{equation}
\boldsymbol{C} = \boldsymbol{L}\boldsymbol{L}^T.
\end{equation}
And then solve the typical linear equation of $\boldsymbol{C}\boldsymbol{G}_\alpha=\boldsymbol{Y}$ from $\boldsymbol{L}$.


For a new input $x_{N+1}$, $\boldsymbol{C}$ can be expanded as follows,
\begin{equation}
    \boldsymbol{C}_{N+1} = 
    \begin{pmatrix}
    \boldsymbol{C}_N & \boldsymbol{k} \\
    \boldsymbol{k}^T & c      
    \end{pmatrix}
\end{equation}
where $c = k(x_{N+1}, x_{N+1}) + \beta$. And the vector $\boldsymbol{k}$ has elements $k(x_n, x_{N+1})$ for $n = 1, \cdots, N$. 

Thus the mean and variance of the new sample can be derived as:
\begin{equation}\label{pred}
    \boldsymbol{Y_{N+1}} = \boldsymbol{C}_{N+1} \boldsymbol{G_\alpha}
\end{equation}
\begin{equation}\label{std}
    \boldsymbol{\sigma}^2_{N+1} = c - \boldsymbol{k}^T \boldsymbol{C}_N^{-1} \boldsymbol{k}
\end{equation}

We can further improve the performance by optimizing $\theta$ and $l$ leading to the maximum log-likelihood (MLL) function. Assuming that the Gaussian noise ($\epsilon_n$) is applied to the observed $\boldsymbol{y_n}$, 
\begin{equation}
    \boldsymbol{t}_n = \boldsymbol{y}_n + \boldsymbol{\epsilon}_n,
\end{equation}

The MLL can be evaluated from the following \cite{bishop2006pattern},
\begin{equation}
    \ln p(\boldsymbol{t}|\theta) = - \frac{1}{2}\bigg[\ln \det(\boldsymbol{C_N}) + \boldsymbol{t}^T\boldsymbol{C_N^{-1}}\boldsymbol{t} + N \ln(2\pi) \bigg] 
\end{equation}

To obtain the best MLL, one can generate multiple trial {$l$} values and then optimize each with the gradient information. In Python programming, this step can be conveniently done by calling \texttt{scipy.optimize} library with the choice of fast optimize like L-BFGS-b \cite{lbfgs-b}.
\begin{equation}
    \frac{\partial}{\partial \theta} \ln p(\boldsymbol{t}|\theta) = 
    - \frac{1}{2} \bigg[\textrm{Tr}\bigg(\boldsymbol{C_N^{-1}} \frac{\partial \boldsymbol{C_N}}{\partial \theta}\bigg) 
    - \boldsymbol{t}^T\boldsymbol{C_N^{-1}} \frac{\partial \boldsymbol{C_N}}{\partial \theta} \boldsymbol{C_N^{-1}} \boldsymbol{t} \bigg]
\end{equation}

\subsection{Structural Descriptor}
Recently, several structural descriptors \cite{BP-PRL-2007, bartok2013on, ACE} have been proposed to characterize the local atomic environment by observing the translation, rotation and permutation symmetry invariance. Following our previous works \cite{zagaceta2020, yanxon2020pyxtalff} , we focus on the use of SO(3) descriptor derived from the power-spectrum of spherical harmonic expansion coefficients. The main idea here is to use the the combined orthonormal radial and angular basis functions to reconstruct the smoothed neighbor density function 
\begin{equation}
    \rho(\mathbf{r}) = \sum_i^{r_i \leq r_c} f_\text{cut}(r_i) e^{(-\alpha|\mathbf{r}-\mathbf{r_i}|^2)}
\end{equation}

where $\alpha$ (typical chosen as 1-2 \AA) is the parameter to control the smoothness of the neighbor density at a given location, and $f_\text{cut}$ is the cutoff function is defined to ensure the density would decay to zero when $r_i$ is approach the $r_c$:

\begin{equation}
    f_{\textrm{cut}}(r) = \begin{cases}
    \frac{1}{2}\left[\cos\left(\frac{\pi r }{r_{\textrm{cut}}}\right) + 1\right],&  r \leq r_{\textrm{cut}}\\
    0,              & r > r_{\textrm{cut}}
    \end{cases}
    \end{equation}

According to Refs. \cite{bartok2013on, zagaceta2020, yanxon2020pyxtalff}, the weighted neighbor density function can be expanded as

\begin{equation}
\rho(\mathbf{r}) = \sum_{nlm} \boldsymbol{c}_{nlm} g_n(r)Y_{lm}(\hat{\mathbf{r}}),
\end{equation}
where $g_n(r)$ is a set of orthonormal radial basis function derived from the polynomials, $Y_{lm}$ is the spherical harmonics, $r$ is the radial distance, $\hat{\mathbf{r}}$ is the unit vector of $\mathbf{r}$. And the expansion coefficients $\boldsymbol{c}_{nlm}$ have three indices to denote the contributions from radial basis ($n$) and the spherical harmonics ($l$ and $m$), respectively.

The analytical expression of $\boldsymbol{c}_{nlm}$ can be derived as follows,

\begin{equation}
\boldsymbol{c}_{nlm} = 4\pi \sum_i^{r_i \leq r_c} f_\text{cut}(r_i) e^{-\alpha r_i^2} Y_{lm}^*(\mathbf{\hat{r}_i}) 
\int_0^{r_c} r^2 g_n(r) e^{-\alpha r^2} I_l(2\alpha r r_i) dr,    
\end{equation}

in which $I_l$ is the modified spherical Bessel function of the first kind.

The $\boldsymbol{c}_{nlm}$ coefficients are complex-valued and rotation variant. To ensure the rotation-invariance, we take its power spectrum combining these expansion coefficients:

\begin{equation}
\boldsymbol{x}_{n_1 n_2 l} = \sum_{m=-l}^{+l}\boldsymbol{c}_{n_1 l m} \boldsymbol{c}^*_{n_2 l m}    
\end{equation}.

Hence, each atom in the structure can be described as an $n_\text{max} \times n_\text{max} \times l_\text{max}$ array. After symmetry reduction, this can be furthered reduced to an 1D array of $n_{\textrm{max}} \times (n_{\textrm{max}}+1) \times (l_{\textrm{max}}+1)/2$. In addition, the derivative of $\boldsymbol{x}$ with respect to all atomic coordinates ($\partial \boldsymbol{x}/\partial \boldsymbol{R}$) will be needed for the computation of forces. The full expression and derivations can be found in Ref. \cite{zagaceta2020}.

For practical applications, we choose $n_\text{max}$= 3, and $l_\text{max}$=4, resulting in a 30-length array to represent a single atomic environment ($\boldsymbol{x}_i$). If a cutoff distance 5-6 \AA~ is chosen, one expects to find 30-50 neighboring atoms, and thus each atom's $\partial \boldsymbol{x}/\partial \boldsymbol{R}$ array is expected to have a size of (30, 50, 3).

\subsection{Gaussian Kernel choices}
After the structure descriptor is known, we define the similarity between two atoms (encoded as $\boldsymbol{x}_1$ and $\boldsymbol{x}_2$) from the cosine distance:
\begin{equation}\label{similarity}
d(\boldsymbol{x}_1, \boldsymbol{x}_2) = \frac{\boldsymbol{x}_1\cdot \boldsymbol{x}_2}{|\boldsymbol{x}_1||\boldsymbol{x}_2|} 
\end{equation},

resulting a scalar between 0 and 1. To improve the distinction capability, one can further apply the power law to get the distance metric ($D$),
\begin{equation}
D(\boldsymbol{x}_1, \boldsymbol{x}_2) = d^{\zeta} 
\end{equation}

where $\zeta$ is a integer to control the distribution.

The corresponding derivatives are:
\begin{equation}
    \begin{split}
        \frac{\partial d}{\partial \boldsymbol{x}_1} &= \frac{\boldsymbol{x}_2 |\boldsymbol{x}_1| - (\boldsymbol{x}_1\cdot \boldsymbol{x}_2) \frac{\boldsymbol{x}_1}{|\boldsymbol{x}_1|}}{|\boldsymbol{x}_2||\boldsymbol{x}_1|^2}\\
        &= \frac{\boldsymbol{x}_2}{|\boldsymbol{x}_1||\boldsymbol{x}_2|} - 
        \frac{(\boldsymbol{x}_1\cdot \boldsymbol{x}_2) \boldsymbol{x}_1}{|\boldsymbol{x}_1|^3|\boldsymbol{x}_2|}
    \end{split}
\end{equation}
\begin{equation}
    \begin{split}
        &\frac{\partial^2 d}{\partial \boldsymbol{x}_1\partial \boldsymbol{x}_2} 
        = \frac{\partial}{\partial \boldsymbol{x}_2} \bigg(\frac{\boldsymbol{x}_2}{|\boldsymbol{x}_1||\boldsymbol{x}_2|} 
        - \frac{(\boldsymbol{x}_1\cdot \boldsymbol{x}_2) \boldsymbol{x}_1}{|\boldsymbol{x}_1|^3|\boldsymbol{x}_2|}\bigg) \\
        &= \frac{\textrm{I}}{|\boldsymbol{x}_1||\boldsymbol{x}_2|} -\frac{\boldsymbol{x}_2\otimes \boldsymbol{x}_2}{|\boldsymbol{x}_1||\boldsymbol{x}_2|^3}  +
        \frac{(\boldsymbol{x}_1 \cdot \boldsymbol{x}_2)\boldsymbol{x}_1 \otimes \boldsymbol{x}_2}{|\boldsymbol{x}_1|^3|\boldsymbol{x}_2|^3} - 
        \frac{\boldsymbol{x}_1 \otimes \boldsymbol{x}_1 }{|\boldsymbol{x}_1|^3|\boldsymbol{x}_2|} \\
    \end{split}
    \end{equation}

    Following the definition of $D$, the derivatives are:
    \begin{equation}
    \begin{split}
        \frac{\partial D}{\partial \boldsymbol{x}_1} &= \zeta d^{\zeta-1}\frac{\partial d}{\partial \boldsymbol{x}_1} \quad
        \frac{\partial D}{\partial \boldsymbol{x}_2} = \zeta d^{\zeta-1}\frac{\partial d}{\partial \boldsymbol{x}_2}\\
        \frac{\partial^2 D}{\partial \boldsymbol{x}_1\partial \boldsymbol{x}_2} 
        &= \frac{\partial}{\partial \boldsymbol{x}_2} \bigg(\frac{\partial D}{\partial \boldsymbol{x}_1}\bigg) 
        = \frac{\partial}{\partial \boldsymbol{x}_2} \bigg(\zeta d^{\zeta-1}\frac{\partial d}{\partial \boldsymbol{x}_1}\bigg) \\
        &= \zeta \Bigg[ d^{\zeta-1} \frac{\partial^2 d}{\partial \boldsymbol{x}_1 \partial \boldsymbol{x}_2} + 
        (\zeta-1) d^{\zeta -2}\frac{\partial d}{\partial \boldsymbol{x}_1} \frac{\partial d}{\partial \boldsymbol{x}_2} \Bigg]\\
    \end{split}
    \end{equation}

After knowing $D$, we can construct two types of kernel functions.

\begin{equation}\label{kernel1}
k(\boldsymbol{x}_1, \boldsymbol{x}_2) = 
\begin{cases}
    \theta^2 \exp[\frac{1}{2}D(\boldsymbol{x}_1, \boldsymbol{x}_2)/l^2] \quad &\mbox{RBF}\\
    \theta^2 [\sigma_0^2 + D(\boldsymbol{x}_1, \boldsymbol{x}_2)]   \quad &\mbox{Dot Product} 
\end{cases}
\end{equation}

For both RBF and Dot Product kernels, we use the following relations to compute the derivatives:
\begin{equation}
\begin{split}
    \frac{\partial k(\boldsymbol{x}_1, \boldsymbol{x}_2)}{\partial \boldsymbol{x}_1} &= \frac{\partial k}{\partial D} \frac{\partial D}{\partial \boldsymbol{x}_1} \\
    \frac{\partial k(\boldsymbol{x}_1, \boldsymbol{x}_2)}{\partial \boldsymbol{x}_2} &= \frac{\partial k}{\partial D} \frac{\partial D}{\partial \boldsymbol{x}_2} \\
\end{split}
\end{equation}

\begin{equation}
    \frac{\partial k}{\partial D} = 
    \begin{cases}
    \frac{0.5\theta^2}{l^2} \exp{(-0.5(1-D)/l^2)} & \mbox{RBF}\\
    \theta^2  & \mbox{Dot Product}\\
    \end{cases}    
    \end{equation}

    \begin{equation}
    \frac{\partial^2 k}{\partial \boldsymbol{x}_1 \partial \boldsymbol{x}_2} = 
    \begin{cases}
    \frac{\partial k}{\partial D}  \bigg[\frac{\partial^2 D}{\partial \boldsymbol{x}_1 \partial \boldsymbol{x}_2} 
    + \frac{1}{2l^2} \frac{\partial D}{\partial \boldsymbol{x}_1}\frac{\partial D}{\partial \boldsymbol{x}_2}\bigg] & \mbox{RBF}\\
    \frac{\partial k}{\partial D}  \frac{\partial ^2 D}{\partial \boldsymbol{x}_1 \partial \boldsymbol{x}_2} & \mbox{Dot Product}\\
    \end{cases}    
\end{equation}

\subsection{Total Energy and Forces}
In a structure $s_1$ with the associated descriptors \{${\boldsymbol{x}^{s_1}, \partial\boldsymbol{x^{s_1}}/\partial \boldsymbol{R}}$\}, its total energy is $E = \sum_1^{N_1} E_i$, where $E_i$ is the linear combination of some basis functions. We can define the summed energy kernel ($k_{\text{ee}}$) to express the covariance of energies between two structures ($s_1$ and $s_2$)

\begin{equation}
    k_{ee}(\boldsymbol{x}^{s_1}, \boldsymbol{x}^{s_2}) = \frac{1}{N_{s_1} N_{s_2}} \sum_{i=1}^{N_{s_1}}\sum_{j=1}^{N_{s_2}}k(\boldsymbol{x}^{s_1}_{i}, \boldsymbol{x}^{s_2}_{j}),
\end{equation}
in which $N_{s_1}$ and $N_{s_2}$ denote the number of atoms in each structure, respectively.

$\boldsymbol{F}$ is related to energy by:
    \begin{equation}
        \boldsymbol{F}_{i} = -\frac{\partial E_{\textrm{total}}}{\partial \boldsymbol{R}_i}  = 
        - \sum_j^{\textrm{all atoms}} \frac{\partial E_j}{\partial \boldsymbol{R}_i} 
    \end{equation}
        
Therefore, we build the derivative kernel ($k_\text{ff}$) between $\boldsymbol{F}_i$ and $\boldsymbol{F}_j$, as long as the descriptor information \{$\boldsymbol{x}_i, \partial \boldsymbol{x}_i/\partial \boldsymbol{R}$\}
and \{$\boldsymbol{x}_j, \partial \boldsymbol{x}_j/\partial \boldsymbol{R}$\} are known.

\begin{equation}
    \begin{split}
    k_\text{ff}(\boldsymbol{x}_i, \boldsymbol{x}_j) 
    &=\sum_{i^\prime}^{N_i} \sum_{j^\prime}^{N_j} k\bigg(
    \frac{\partial E_i}{\partial \boldsymbol{x}_{i^\prime}}
    \frac{\partial \boldsymbol{x}_{i^\prime}}{\partial \boldsymbol{R}_i},
    \frac{\partial E_j}{\partial \boldsymbol{x}_{j^\prime}} 
    \frac{\partial \boldsymbol{x}_{j^\prime}}{\partial \boldsymbol{R}_j} \bigg) \\
    &= \sum_{i^\prime}^{N_i} \sum_{j^\prime}^{N_j} 
    \bigg( \frac{\partial \boldsymbol{x}_{i^\prime}}{\partial \boldsymbol{R}_i} \bigg)^T 
    \frac{\partial^2 k(\boldsymbol{x}_{i^\prime}, \boldsymbol{x}_{j^\prime})}{\partial \boldsymbol{x}_{i^\prime} \partial \boldsymbol{x}_{j^\prime}} 
    \frac{\partial \boldsymbol{x}_{j^\prime}}{\partial \boldsymbol{R}_j}, \\
    \end{split}
\end{equation}
where $N_i$ and $N_j$ denote the number of neighbors within the cutoff distances.

Next, we also need the kernel ($k_\text{ef}$) between energy $E_{s_1}$ and $\boldsymbol{F}_j$
\begin{equation}
  k_\text{ef}(\boldsymbol{x}^{s_1}, \boldsymbol{x}_j) = k\bigg(E_{s_1}, \frac{\partial E_j}{\partial \boldsymbol{R}_j}\bigg)
= \frac{1}{N_{s_1}}\sum_{i=1}^{N_{s_1}} \sum_{j^\prime=1}^{N_j} \frac{\partial k(\boldsymbol{x}_i, \boldsymbol{x}_{j^\prime})}{\partial \boldsymbol{x}_{j^\prime}}\frac{\partial \boldsymbol{x}_{j^\prime}}{\partial \boldsymbol{R}_j}
\end{equation}

Assuming there exists $N_e$ energy points ($\boldsymbol{x}^{s_1}, \boldsymbol{x}^{s_2}, \dots)$ and $N_f$ force points ($\boldsymbol{x}^0_1, \boldsymbol{x}^0_2 $), the total GPR covariance is a combination of energy/force kernels
\begin{equation*}
\boldsymbol{C}_{\textrm{total}} = 
\begin{pmatrix}
\boldsymbol{K}_\text{ee} & \boldsymbol{K}_\text{ef} \\
\boldsymbol{K}_\text{ef}^T & \boldsymbol{K}_\text{ff} 
\end{pmatrix}
+ \beta \boldsymbol{\delta}_{mn}
\end{equation*}
which is a $N_c \times N_c$ square matrix ($N_c = N_e + 3~\times~ N_f$).

After knowing the covariance matrix $\boldsymbol{C}_{\textrm{total}}$, the $\boldsymbol{G_\alpha}$ can be determined.

To predict the energy and forces for a new structure $\{\boldsymbol{x}^t, \boldsymbol{\partial \boldsymbol{x}^t/\partial{R}}\}$, one can follow eq. \ref{pred} to get the energy and force predictions.

\begin{equation}
    E^t = \begin{bmatrix} k_\text{ee}(\boldsymbol{x}^{t}, \boldsymbol{x}^{s_1}) \\ \vdots \\ k_\text{ee}(\boldsymbol{x}^{t}, \boldsymbol{x}^{s_n}) \\ 
    k_\text{ef}(\boldsymbol{x}^{t}, \boldsymbol{x}_1) \\ \vdots \\ k_\text{ef}(\boldsymbol{x}_{t_1}, \boldsymbol{x}^0_{3n}) \end{bmatrix} \cdot \boldsymbol{G}_\alpha
\quad
    \boldsymbol{F}^t_i = \begin{bmatrix} k_\text{fe}(\boldsymbol{x}_i, \boldsymbol{x}_{s_1}) \\ \vdots \\ 
    k_\text{fe}(\boldsymbol{x}_i, \boldsymbol{x}_{ss}) \\ k_\text{ff}(\boldsymbol{x}_i, \boldsymbol{x}^0_{1}) \\ \vdots \\ 
    k_\text{ff}(\boldsymbol{x}_i, \boldsymbol{x}^0_{3n}) \end{bmatrix}
    \cdot \boldsymbol{G}_\alpha
\end{equation}

\begin{figure*}[htbp]
    \centering
    \includegraphics[width=0.9\linewidth]{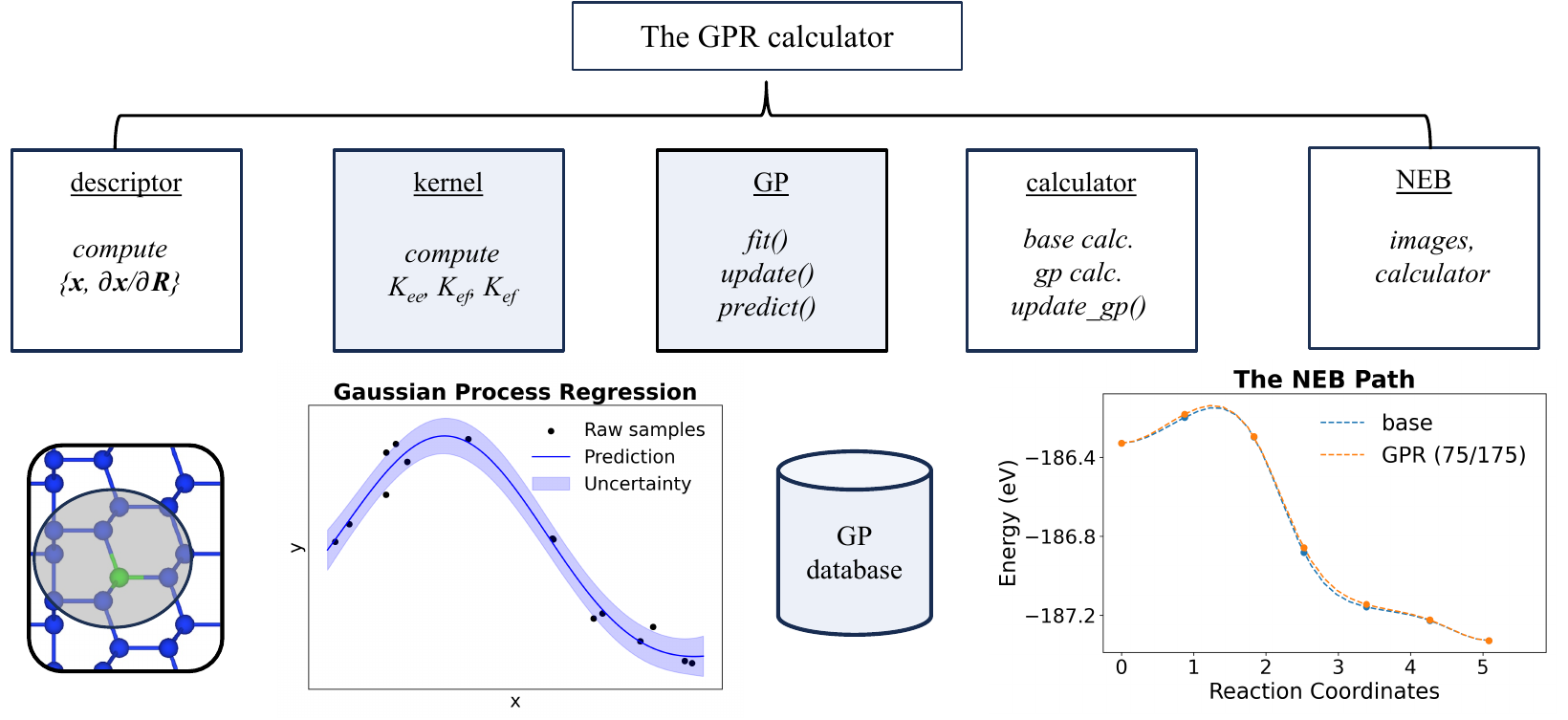}
    \caption{The list of available modules in the \texttt{GPR\_calculator} package and their corresponding outputs. The modules with accerlation are also highlighted in light blue boxes.}
    \label{fig1}
\end{figure*}

\section{Algorithm Implementation}
To enable the GPR framework for practical NEB calculation, we have developed different modules to handle descriptor calculation, GPR regression, hybrid calculator implementation, and interfaces with existing NEB calculation tools. We also incorporate parallelization to enable fast computation.

 \subsection{Core Modules}
The package consists of five main modules that handle different aspects of the calculations, as summarized in Figure \ref{fig1}:

\begin{enumerate}
    \item \texttt{gpr\_calc.descriptor}: Processes input structures and computes structural descriptors.
    \item \texttt{gpr\_calc.kernel}: Initializes RBF and Dot Product kernels with predefined parameters, computes different kernel functions ($k_\text{ee}$, $k_\text{ef}$, $k_\text{ff}$), and constructs the total covariance matrix $C_\text{total}$.
    \item \texttt{gpr\_calc.GP}: Provides the framework for optimizing kernel parameters, predicts energies for new structures, and enables dynamic updates of the training dataset and covariance matrix.
    \item \texttt{gpr\_calc.calculator}: Implements the hybrid calculator that combines base electronic structure calculations with on-the-fly GPR predictions based on uncertainty thresholds.
    \item \texttt{gpr\_calc.NEB}: Contains utilities and interfaces with \texttt{ASE}'s \texttt{ase.mep.NEB} module, including functions for initializing images, running NEB calculations, and plotting the results.
\end{enumerate}

\subsection{Acceleration and Parallelization}

Compared to other machine learning models such as neural networks, GPR is known to suffer from scalability issues, with computational costs scaling cubically with dataset size. In our application, additional expensive components include the computation of descriptors and kernels, particularly for $k_\text{ff}$, $k_\text{ef}$ and $k_\text{fe}$ computations. For large numbers of structures, descriptor computation can also become a bottleneck, especially in Python implementations.

To address these challenges while maintaining Python's flexibility and ease of use, we have implemented several optimization strategies:

First, we rewrote the kernel computation in C++ and used the \texttt{cffi} library to interface with the Python code. This allows us to leverage C++'s computational speed while preserving Python's flexibility and rich ecosystem.

Second, we parallelized the GPR calculator using the \texttt{mpi4py} library \cite{rogowski2022mpi4py}. In the GPR calculation, each process computes the covariance matrix for a subset of the data points. This enables efficient scaling to large datasets by distributing computations across multiple processors. When a large training database is loaded, the computation of structural descriptors is also parallelized across multiple processes.

\subsection{The overall NEB-GPR workflow}

\begin{figure*}[htbp]
    \centering
    \includegraphics[width=0.9\linewidth]{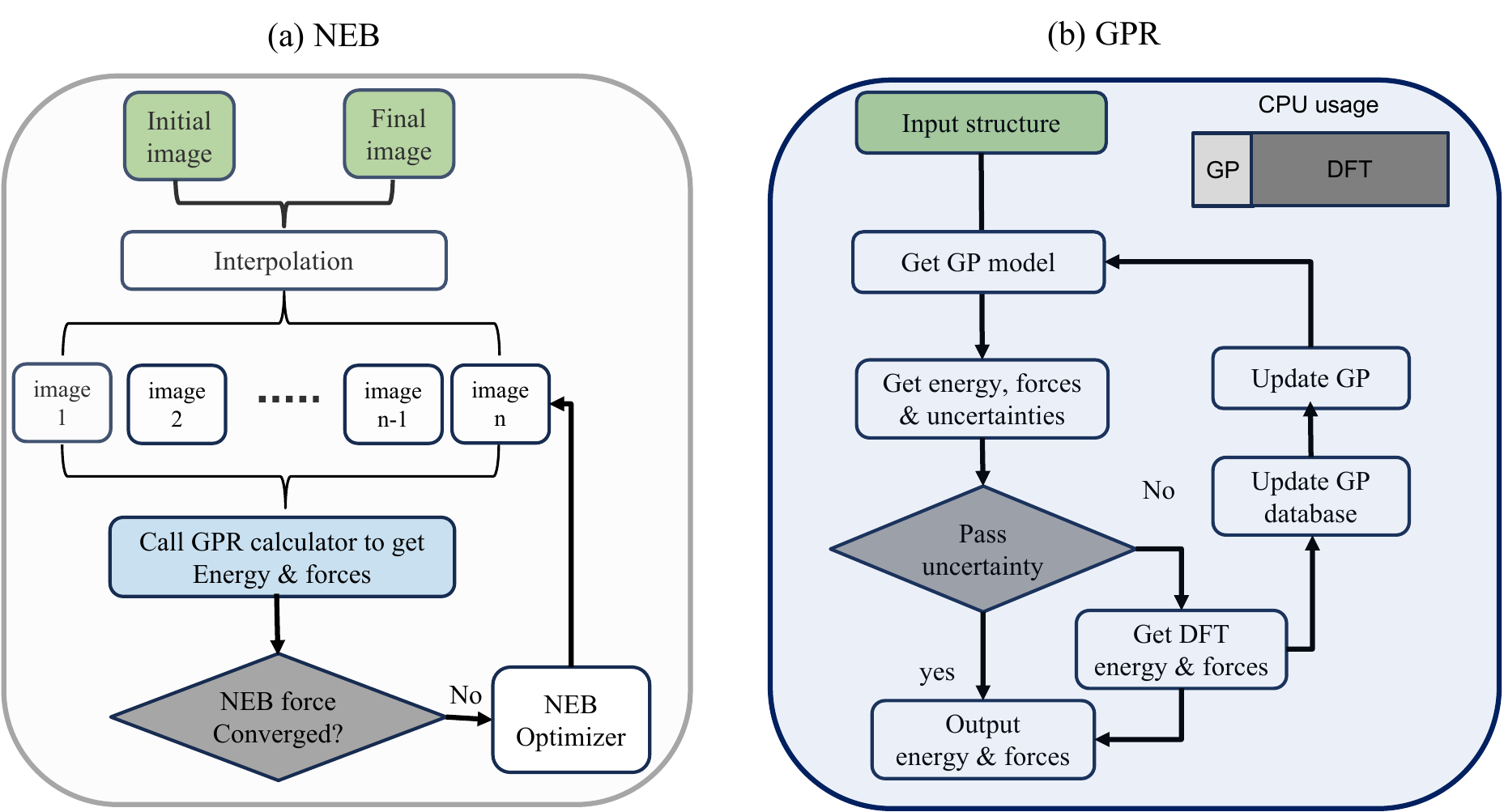}
    \caption{The workflow of NEB-GPR calculation. (a) In a NEB run, each image's energy and forces are predicted by the GPR calculator until the NEB calculation converges. (b) The GPR calculator use either GP model or the base calculator to yield the energy and forces. If the base calculator is called, the GP model is updated with the new image's energy and forces.}
    \label{fig2}
\end{figure*}

Figure \ref{fig2} shows the overall workflow for a practical NEB calculation using the GPR calculator. The NEB-GPR workflow consists of two main steps: (1) the NEB module handles the input images to generate initial trajectory and then optimize the trajectory based on the calculated energy and forces for each image; (2) the GPR calculator predicts the energy and forces for each image in the NEB calculation. If the GPR calculator is uncertain about its predictions, it calls the base calculator to compute the energy and forces for that image. The GPR model is then updated with the new image's energy and forces. This process continues until the NEB calculation converges.

To ensure the GPR calculator is not overly reliant on the base calculator, we set a threshold for the maximum uncertainty in energy and forces (e.g. $\sigma_E$ = 0.05 eV/structure for energy and $\sigma_F$=0.075 eV/\AA~ for forces). If the predicted uncertainty exceeds this threshold, the GPR calculator will call the base calculator to compute the energy and forces for that image. This allows us to balance the computational efficiency of the GPR calculator with the accuracy of the base calculator. Since a practical NEB calculation mainly seeks to minimize the spring forces, we choose to use only $\sigma_F$. For each single point energy calculation, the GPR calculator will check the predicted forces and compare them with the threshold ($\sigma_F$). If the predicted forces are larger than the threshold and one third of the predicted forces, the GPR calculator will call the base calculator to compute the energy and forces for that image. The DFT energy, as well as the atomic forces with large uncertainty, will be then added to the GP database and then used to update the GP model. This allows us to ensure that the NEB calculation converges to a physically meaningful result.

For NEB calculations on supercomputer nodes, we partition the available physical cores into two groups: one for the GPR calculator and the other for the base calculator. The GPR calculator is responsible for computing the covariance matrix and predicting energies, while the base calculator handles the actual electronic structure calculations. This division of labor allows us to fully utilize the available computational resources and achieve significant speedup in NEB calculations. To avoid frequent GP model update, we usually check if there is a need to update the model at the end of each NEB iteration.

\section{Dependencies}
\label{dependence}
The majority of codes are written in Python 3.9 or higher. Like many other Python packages, it relies on several external Python libraries. This library is intended to interface with \texttt{ASE} for practical calculations and hence it inherits all package dependencies (e.g., Numpy \cite{numpy}, Scipy \cite{scipy}) required by ASE. 

In addition, the cffi \cite{cffi} and mpi4py \cite{rogowski2022mpi4py} libraries are required for the C++ kernel implementation and parallelization, respectively. As such, the package requires a C++ compiler and an OpenMPI implementation to be installed on the system. The package is compatible with both Linux and macOS operating systems.

\section{Installation and Example Usages}
\label{usage}
The \texttt{GPR\_calculator} package can be installed using pip. The package is compatible with Python 3.9 and later versions. To install, simply download the source code from \url{https://github.com/MaterSim/GPR_calculator}, change directory to the root folder and then run \texttt{pip install .} from the terminal.

The following simple example demonstrates how to set up a GPR calculator, perform a NEB calculation, and visualize the results.

\begin{lstlisting}[language=Python, caption=A Python script to perform a NEB calculation using GPR\_calculator]
from ase.calculators.emt import EMT
from gpr_calc.gaussianprocess import GP
from gpr_calc.calculator import GPR
from gpr_calc.NEB import neb_calc, get_images, plot_path
    
# Set parameters
init = 'database/initial.traj'
final = 'database/final.traj'
fmax = 0.05
    
# Run NEB with EMT calculator
images = get_images(init, final, 5)
neb = neb_calc(images, EMT(), fmax=fmax)
label = f'EMT ({steps*(len(imgs)-2)+2})'
dat = [(neb.images, neb.energies, label)]
    
# Run NEB with gpr calculators
for tol in [0.05, 0.1]:
    imgs = get_images(init, final, 5)
    
    # Initialize GPR model & calculator
    noise_e = tol / len(images[0])
    gp = GP.set_GPR(images, EMT(),
                    noise_e=noise_e,
                    noise_f=tol)
    calc = GPR(base=EMT(), ff=gp)
    
    # Run NEB calculation
    neb = neb_calc(images, calc, fmax=fmax)
    N1 = gp.use_base
    N2 = gp.use_surrogate
    label = f'GPR-{tol:.2f} ({N1}/{N2})'
    dat.append((neb.images, neb.energies, label))
    
# Plot the results
neb_plot_path(dat, figname='path.png', 
              title='Au diffusion on Al(100)')
\end{lstlisting}

In this example, we first import the necessary modules and set the initial and final structures, as well as the maximum force tolerance. We then run the NEB calculation using the effective medium theory (EMT) calculator \cite{EMT} and store the results in the \texttt{dat} list. Next, we run the NEB calculation using the GPR calculator for different energy/force tolerances and store the results in the \texttt{data} list. Finally, we plot the results using the \texttt{plot\_path} function. 

\begin{figure}[h]
    \centering
    \includegraphics[width=0.46\textwidth]{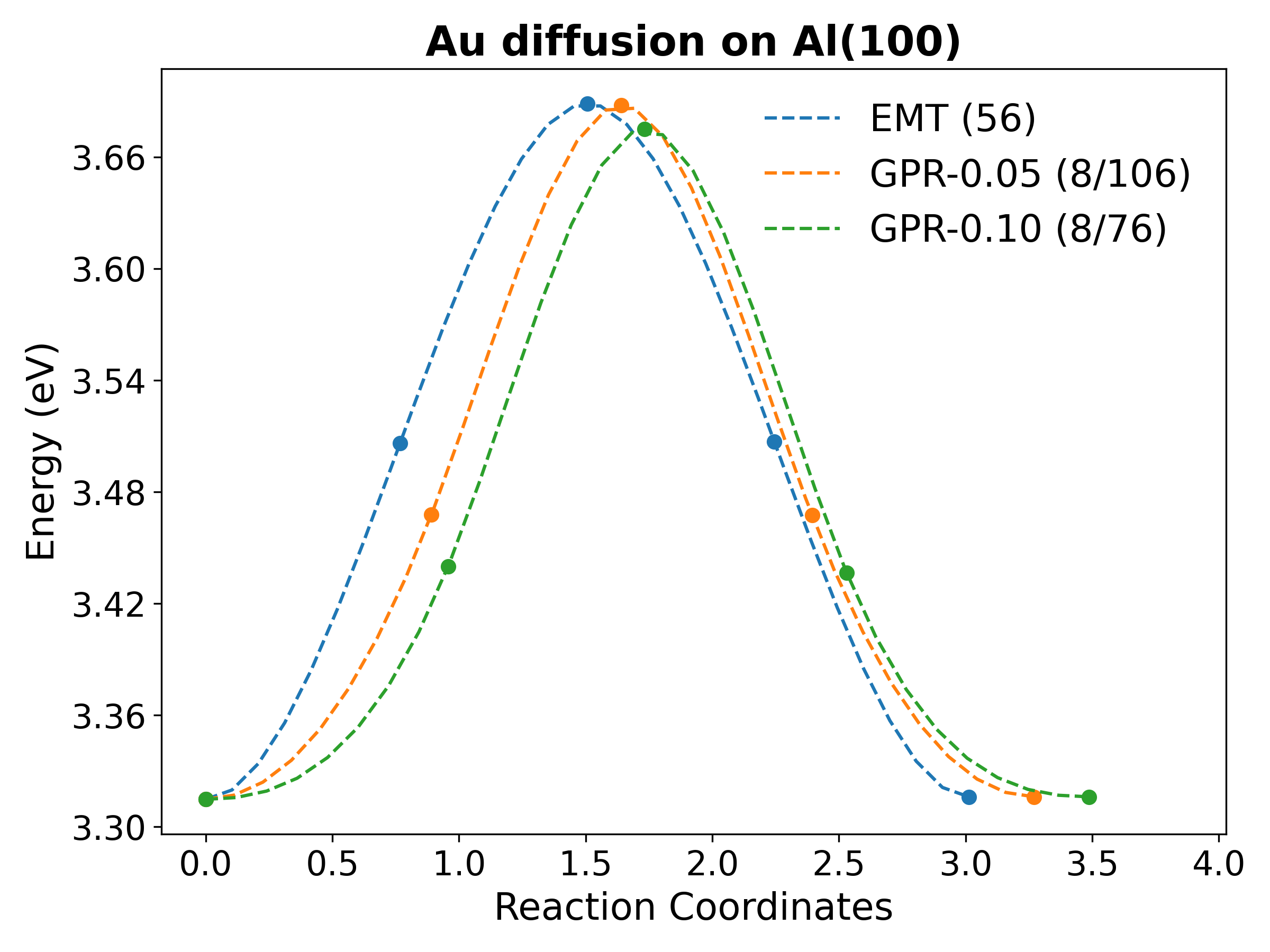}
    \caption{The simulated Au diffusion on Al(100) surface from the EMT and GPR calculators with different energy tolerances. The number of EMT/GPR calculations required for convergence is shown in parentheses.}
    \label{fig:neb_plot}
\end{figure}

The resulting plot in Figure \ref{fig:neb_plot} illustrates the energy profile of Au diffusion on the Al(100) surface, comparing the EMT and GPR calculators. Notably, the GPR calculator with a smaller tolerance setting (0.05 eV for energy and 0.05 eV/\AA~ for forces) produces results close to the EMT-driven NEB calculation, but requires only 8 EMT single-point energy evaluations, compared to 53 for the pure EMT calculation. Conversely, the GPR calculator with a larger tolerance setting (0.10 eV for energy and 0.10 eV/\AA~ for forces) exhibits a more noticeable deviation.

Due to the existence of uncertainty, it is common that surrogate-driven NEB simulation requiring more iterations to converge to the desired minimum. Despite this, the GPR calculator still substantially reduces the computational effort needed for convergence. Given that DFT calculations, which are more accurate and expensive than EMT, are typically used as the base calculator, a significant speedup (3-10 times) is anticipated. For optimal performance, users can adjust the tolerance values to balance accuracy and computational efficiency.

\section{Practical Applications}
\label{examples}
To demonstrate the capabilities of the \texttt{GPR\_calculator} package, we focus on the use of GPR models to accelerate NEB calculations for surface diffusion and dissociation reactions, relative to traditional DFT-driven approaches. The examples presented here are intended to showcase the versatility and efficiency of the package in a variety of scenarios.

In both examples, our DFT calculations are performed using the Vienna \textit{ab initio} simulation package (VASP) \cite{VASP01-1996, VASP02-1996, VASP03-1993}. The core electrons and the electron-electron exchange correlation effects are treated using the projector augmented wave (PAW) \cite{Bloch-1994, PAW-potentials-1999} method and the Perdew-Burke-Ernzerhof (PBE) functional \cite{PBE-PRL-1996}, respectively. The valence electrons are modeled using a plane-wave basis set expanded to a cutoff energy of 400 eV. The Methfessel-Paxton smearing method \cite{Methfesel-1989} with a smearing width of 0.1 eV, and the calculations are deemed to have converged with energy and force tolerances of $10^{-4}$ eV and 0.03 eV/\AA~, respectively. For NEB optimization, the calculations were considered to have converged when the force on the images is less than 0.075 eV/\AA~ using the FIRE algorithm \cite{FIRE-2006}. The image dependent pair potential (IDPP) method \cite{IDPP-2014} is preferred over linear interpolation to generate the initial guess of the MEP because it produces a more physically meaningful pathway. IDPP prevents atoms from being placed too close to each other, which would otherwise lead to large energy and force spikes. This helps avoid potential divergence in electronic structure calculations. 

In the following examples, we run each system with a single compute node of AMD EPYC 9654P 96-Core Processor with 2.40 GHz. For GPR calculations, we set up the GPR calculator with a noise of 0.05 eV/structure for energy and 0.075 eV/\AA~ for forces, a cutoff distance of 5.0 \AA~ for the descriptor calculation, and the energy and forces are predicted using the RBF kernel. Out of 96 cores, 24 cores are used for the GPR calculator, while the remaining 72 cores are used for the base calculator. For the pure VASP calculations, all 96 cores were used for each single point energy calculation of the NEB image. 

\subsection{Pd\texorpdfstring{$_4$}{4} cluster diffusion on MgO(100) surface}

We first consider the diffusion of a Pd$_4$ cluster on the MgO(100) surface, which is a classical NEB example as discussed in the previous literature \cite{Henkelman-2008}. The initial state consists of a Pd$_4$ cluster adsorbed on the MgO(100) surface, while the final state features the same cluster in a different position on the surface. MgO(100) is modeled as a three-layer surface (72 atoms per layer) with atoms in the top two layers relaxed and those in the bottom two layers fixed in their bulk positions, consistent with Henkelman \textit{et. al.} \cite{Henkelman-2008}. A 15 \AA~ vacuum is included and the Brillouin zone (BZ) is sampled using a $(2\times 2 \times 1)$ Monkhorst-Pack k-point mesh. 

\begin{figure}[h]
    \centering
    \includegraphics[width=0.48\textwidth]{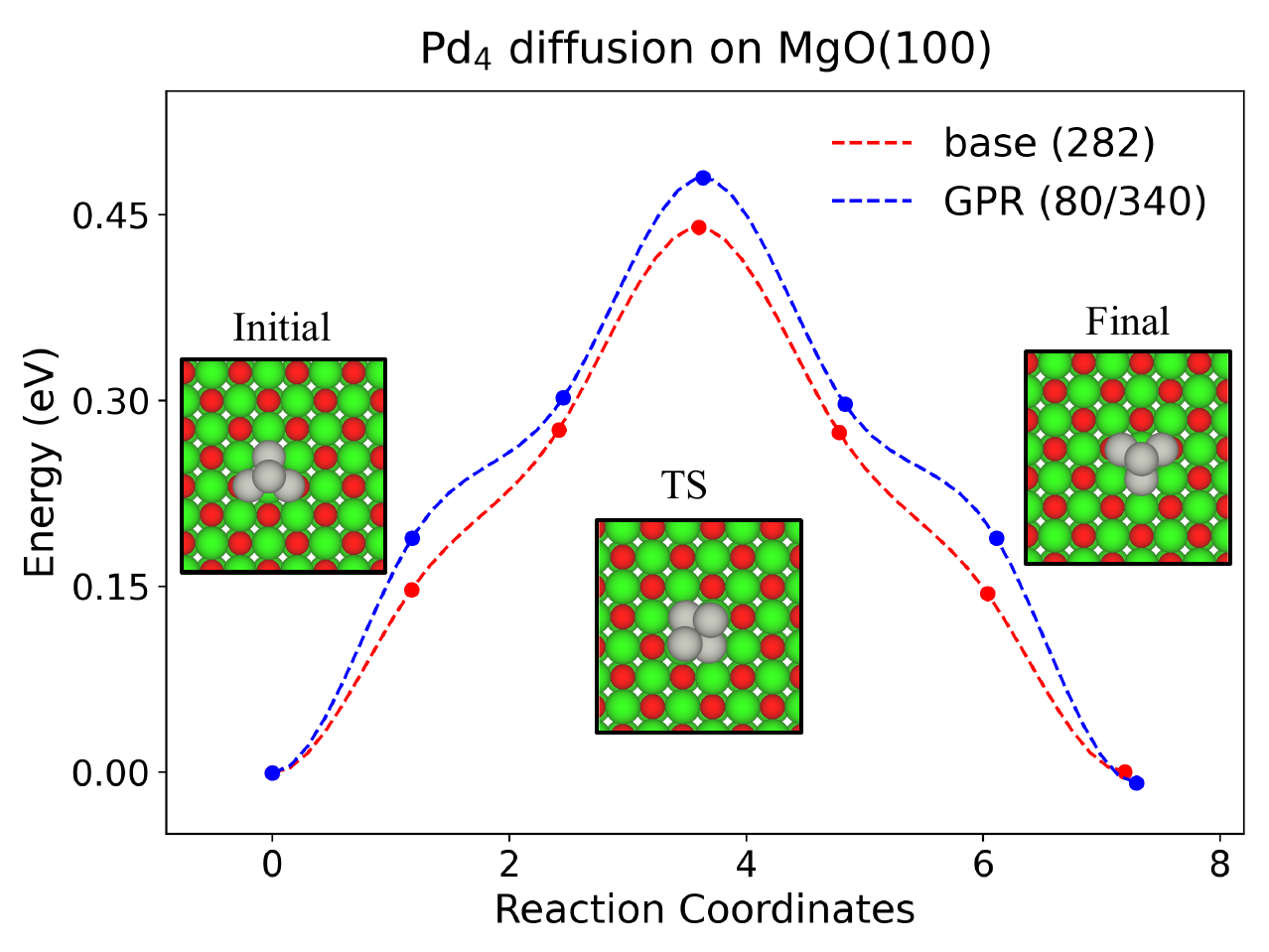}
    \caption{The simulated MEP of Pd$_4$'s dissociation on the MgO(100) surface from both the GPR and pure VASP calculators. The representative structures along the transition path are also shown in the inset.}
    \label{fig4}
\end{figure}

Figure \ref{fig4} shows the MEP for the diffusion of the Pd$_4$ cluster on the MgO(100) surface, as calculated using both the GPR and pure VASP calculators. Clearly, both calculators yield nearly identical MEPs, with the GPR calculator producing a slightly higher activation barrier of 0.48 eV compared to 0.44 eV from the pure VASP calculator. Despite the energy difference, the transition states (TS), featured as a rolling of Pd$_4$ to an adjacent site (as shown in Figure \ref{fig4}), are nearly identical in both cases.

Remarkably, the GPR calculator the demonstrates a substantial acceleration, completing the MEP optimization within 6 hours, consisting of 80 VASP calls and 340 GPR calls. As compared to the VASP results (36.5 hours consisting of 337 DFT calls), the GPR achieves resulting a 5 times speed-up. The GPR calculator's efficiency is attributed to its ability to predict energies and forces on-the-fly. In fact, the majority of NEB simulation can be achieved with less than 50 VASP calls and 110 GPR calls when the maximum NEB force is 0.09 eV/\AA. The rest of simulation were mostly used to refine the final results. 

For a typical  diffusion energy barrier less than 0.50 eV, the GPR can achieve satisfactory accuracy with a small number of VASP calls. This example highlights the potential of the GPR calculator to accelerate NEB calculations while maintaining accuracy.

\subsection{H\texorpdfstring{$_2$}{2}S dissociation on Pd(100) surface}

To test the feasibility of the GPR calculator for chemical reactions, we consider a more challenging system: the dissociation of H$_2$S on a Pd(100) surface. This system is not only of great interest in catalysis and surface chemistry, but also computationally more challenging as it involves the breaking of chemical bonds and the formation of new ones \cite{Alfonso-2005, Alfonso-2008}. The initial state consists of a single H$_2$S molecule adsorbed on the Pd(100) surface, while the final state features two H atoms and an S atom adsorbed on the surface. The Pd(100) surface is modeled as a four-layer slab (9 atoms per layer) with the top two layers relaxed and the bottom two layers fixed in their bulk positions. A 15 \AA~ vacuum is included and the Brillouin zone (BZ) is sampled using a $(3\times 3 \times 1)$ Monkhorst-Pack k-point mesh. The NEB calculations for this system are considered to have converged when the maximum force on each image is less than 0.075 eV/\AA~. To ensure the saddle point is correctly identified along the NEB path, we used the climbing image method \cite{CINEB-2000} instead of the regular NEB method.

\begin{figure*}[htbp]
    \centering
    \includegraphics[width=0.9\linewidth]{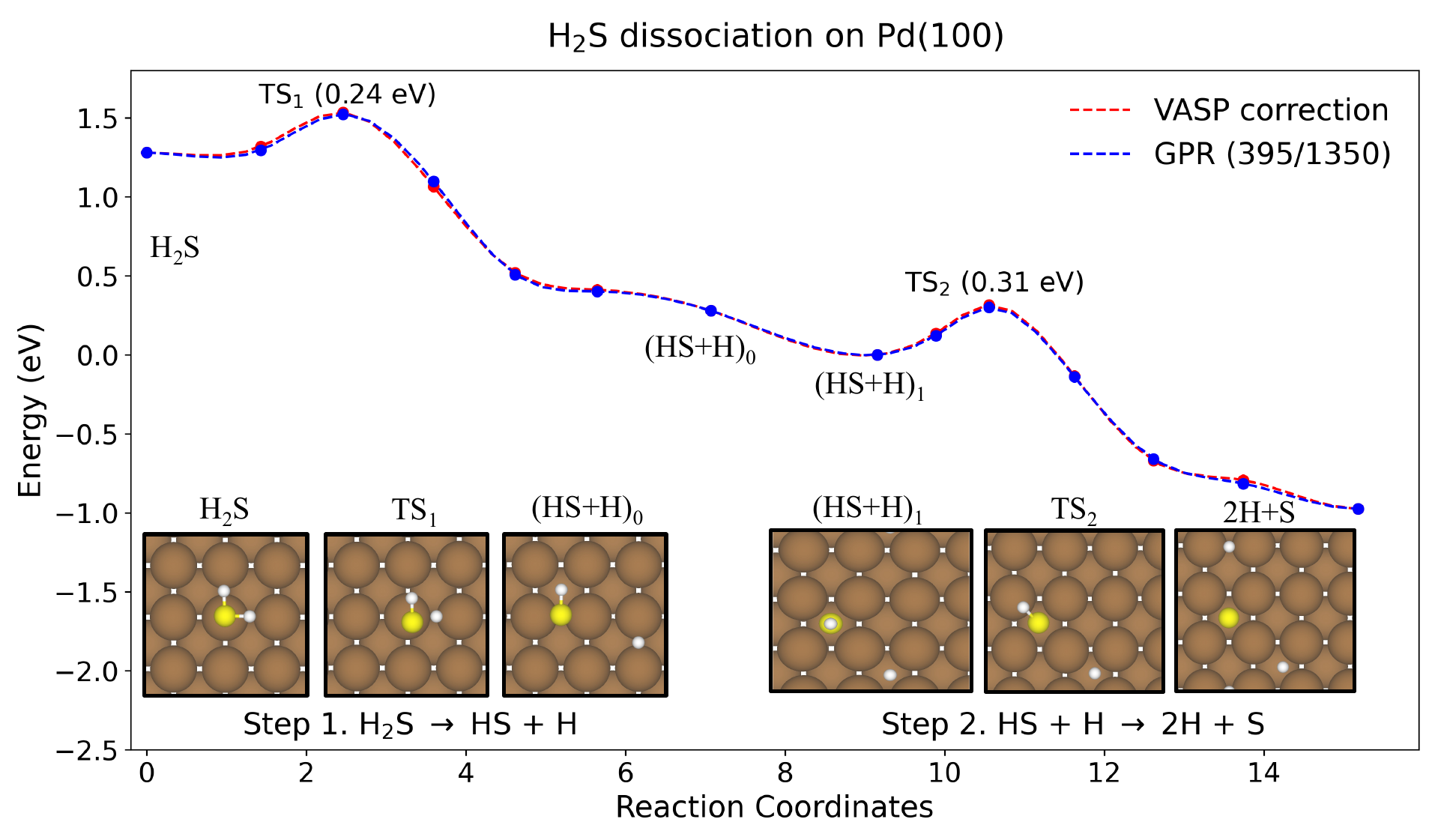}
    \caption{The simulated MEP of H$_2$S's dissociation on the Pd(100) surface from both the GPR and pure VASP calculators. The representative structures along the transition path are also shown in the inset.}
    \label{fig5}
\end{figure*}

Figure \ref{fig5} shows the MEP of H$_2$S dissociation on the Pd(100) surface, calculated using the GPR calculator. For comparison, the final NEB trajectories are also recomputed with the pure VASP calculator. Clearly, the energy profiles from GPR and VASP are nearly identical (with no more than 0.015 eV for each image), confirming the accuracy of GPR model. Due to the inherent uncertainty in GPR models, we found that convergence becomes challenging when the maximum NEB forces approach 0.075 eV/\AA. To validate the reliability of this GPR-optimized NEB trajectory, we perfomed a NEB calculation using a pure VASP calculator the GPR-generated path as the initial MEP guess. After 25 NEB iterations, the trajectory successfully converged to a maximum NEB force below 0.05 eV/\AA, exhibiting only minimal structural changes. 

According to Figure \ref{fig5}, the dissociation occurs in two steps with activation barriers of 0.24 eV and 0.31 eV, respectively. The mechanism involves sequential H-S bond breaking: first H$_2$S $\rightarrow$ HS + H, followed by HS + H $\rightarrow$ 2H + S. In the initial state, the H-S bond length is 1.34 \AA. During dissociation, both H-S bonds gradually elongate as the hydrogen atoms separate from the sulfur. In the first step, the dissociated H-S and H fragments from H$_2$S initially on the top site migrate to adjacent bridge sites, with the H-S distance extending to 1.88 \AA~ in its transition state (TS$_1$). In the second step, the H-S and H fragments begin at hollow sites, their most favorable adsorption positions. The H-S fragment first rotates, after which the S atom remains at the hollow site while the H migrates to the top of neighboring Pd site, causing the H-S length to increase from 1.33 \AA~ to 1.42 \AA~ in TS$_2$. Beyond TS$_2$, the H atom follows a curved path to reach its final bridge site position.


In terms of computational efficiency, the GPR calculator completes the whole MEP optimization in 4.5 hours with 395 VASP calls and 1350 GPR calls. Given that the GPR cost is much lower than that of VASP computations, this approximately represents a 3.5 times speedup. This acceleration is slightly less than in the previous example due to the more complex and asymmetric energy surface requiring more VASP calls.

This example highlights the \texttt{GPR\_calculator}'s capability to handle complex chemical reactions, while maintaining accuracy and efficiency. When the energy profile is more complex and it requires more NEB iterations to reach the converged pathway, the surrogate model is expected to result in a better speed up.
In computational catalysis design, the ability to accurately predict transition states and reaction pathways is crucial for understanding catalytic mechanisms. The \texttt{GPR\_calculator}'s performance in this example demonstrates its potential as a valuable tool for studying complex chemical processes.

\subsection{H\texorpdfstring{$_2$}{2}S dissociation on (100) transion metal surfaces}

\begin{figure*}[h]
    \centering
    \includegraphics[width=0.9\textwidth]{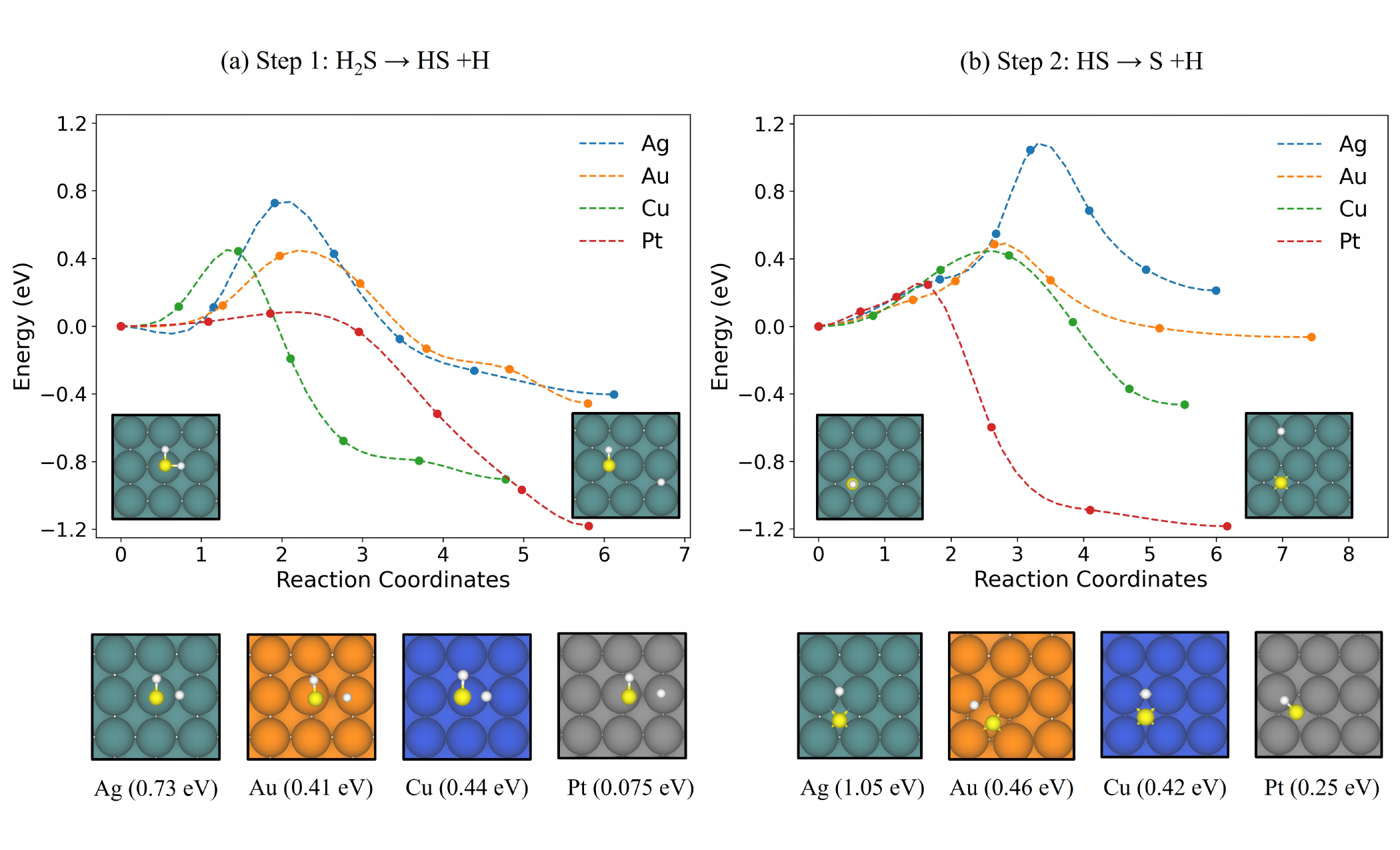}
    \vspace{-5mm}
    \caption{The simulated MEP of H$_2$S dissociation on the (100) surfaces using the GPR calculator for (a) H$_2$S $\rightarrow$ HS + H; and (b) HS $\rightarrow$ S + H. In the lower panel, the transition states for each step are also shown. Green, orange, blue, gray, yellow, and white spheres represent Ag, Au, Cu, Pt, S and H atoms, respectively.}
    \label{fig6}
\end{figure*}

Having demonstrated the capabilities of the GPR calculator for modeling H$_2$S dissociation on Pd, we extend its application to other (100) transition metal surfaces: Au, Ag, Cu, and Pt. The computational setup is similar to that of the Pd example above, with stricter criterion that NEB calculations are considered converged when the maximum force on each image falls below 0.05 eV/\AA~ and a $(2\times 2 \times 1)$ k-point mesh. The whole results, including both energy profile and transition state configurations, are summarized in Figure \ref{fig6}.

The first step of dissociation, as shown in Figure \ref{fig6}a, is analogous to that on Pd, where H-S and H dissociated fragments from the H$_2$S adsorbed on a top site, migrate to adjacent bridge sites. At the transition state, the H fragment is closer to the top site for Pt, whereas it on the bridge site for Au, Ag and Cu. Additionally, the distances between S atom and the migrating H atom at the transition state increase from 1.36 \AA~ to 2.37 \AA~, 2.12 \AA~, 2.17 \AA~, and 1.75 \AA~ for Au, Ag, Pt, and Cu, respectively. Among these, Pt exhibits the most favorable energetics for hydrogen abstraction, with a low reaction barrier of 0.075 eV and a highly exothermic reaction energy of -1.18 eV, indicating facile and strongly favorable dissociation. Cu and Au also facilitate this step, with reaction energies of -0.91 eV and -0.46 eV, and moderate barriers of 0.44 eV and 0.41 eV, respectively. Ag is the least reactive surface, with reaction energy of -0.40 eV and the highest barrier at 0.73 eV.

In the second dissociation step (Figure \ref{fig6}b), the H-S fragment begins at a hollow site, with the sulfur atom remaining at that position while the hydrogen migrates to a bridge site. The reaction path is approximately linear on Ag and Cu but appears curved on Au and Pt, as observed on Pd. The S-H distances at the transition state increase from 1.37 \AA~ to 1.83 \AA~, 2.13 \AA~, 1.56 \AA~, and 1.40 \AA~ for Au, Ag, Cu, and Pt, respectively. This step is energetically unfavorable on Ag and Au, with Ag being endothermic (0.21 eV) and Au nearly thermoneutral (-0.07 eV), accompanied by barriers of 1.05 eV and 0.46 eV, respectively. Conversely, the reaction is exothermic on Cu(-0.46 eV) and Pt(-1.18 eV), with low barriers of 0.42 eV and 0.25 eV, respectively. As in the first step, Pt(100) exhibits the most favorable energetics, combining the lowest activation energy with the most exothermic reaction profile. 

Overall, our simulation results suggest that the H$_2$S dissociation is less favorable on noble metals (Ag, Au) than on transition metals (Cu, Pd, Pt). Both Pd and Pt surfaces readily facilitate dissociation, characterized by low energy barriers and strongly exothermic reaction profiles. These trends align with previous DFT studies by Alfonso et al. \cite{Alfonso-2008}, confirming the accuracy of the GPR calculator. 

\begin{table}[h]
    \centering
    \caption{The detailed GPR/VASP usage and timing for H$_2$S dissociation on different (100) metal surfaces on a single AMD EPYC 9654P 96-Core code.}
    \label{tab:summary}
    \begin{tabular}{ccccc}
        \hline\hline
        Surface & VASP calls & GPR calls & Total time (hours) \\
        \hline
        Ag(100) & 513 & 1912 & 4.60 \\
        Au(100) & 691 & 1304 & 9.55 \\
        Cu(100) & 317 & 1133 & 2.75 \\
        Pt(100) & 683 & 2207 & 14.43 \\
        \hline\hline
    \end{tabular}
\end{table}

Table \ref{tab:summary} summarizes the breakdowns for all calculations. Generally, the number of GPR evaluations exceeded that of VASP by at least a factor of three, with the exception of the Au(100) surface, which required slightly less than 2 times more GPR calls. This deviation is likely stems from a more intricate potential energy surface associated with Au(100), characterized by more substantial atomic displacements of the surface atoms during relaxation, especially pronounced in the second dissociation step. These larger structural rearrangements make it more challenging for the GPR model to accurately interpolate and predict forces across configuration space. Overall, we anticipate that the GPR calculator achieves a speedup of 3-5 times compared to pure VASP calculations.

\section{Remarks on the Limitation and Potential Serendipity}
As demonstrated in the previous sections, surrogate models often require more iterations to converge. When the system approaches the minimum energy pathway, high force accuracy becomes crucial, necessitating frequent calls to the base calculator and updates to the GP model. This can lead to unnecessary optimization steps in NEB iteration. To prevent such inefficient exploration, it is recommended to set a slightly larger force tolerance (e.g., 0.075 - 0.100 eV/\AA~). For barrier calculations, we found that this approach typically yields activation energies within 0.05 eV error of the true value. When DFT-level accuracy is required, one can perform a quick pure DFT-NEB calculation using the GP-optimized trajectory as the initial guess, which typically converges rapidly.

\begin{figure}[h]
    \centering
    \includegraphics[width=0.48\textwidth]{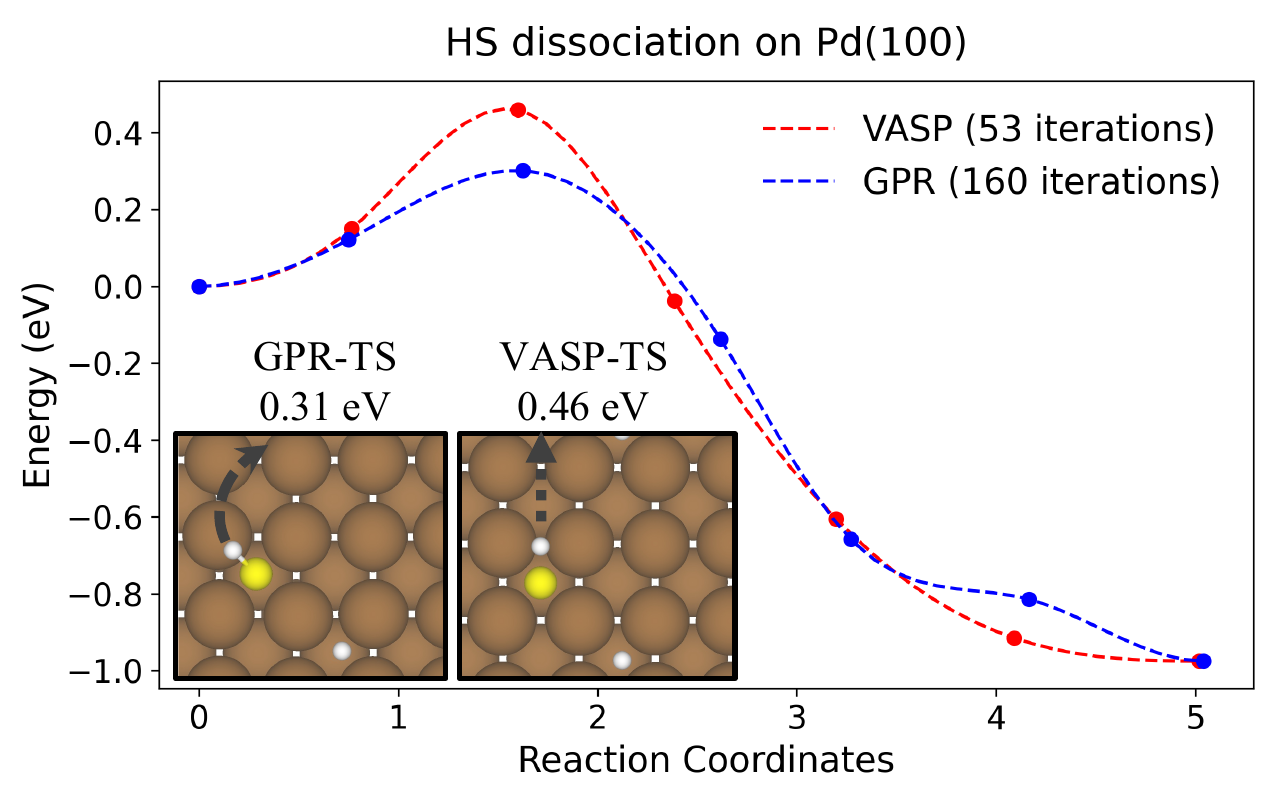}
    \caption{The simulated MEP of HS dissociation on the Pd(100) surface from both the GPR and pure VASP calculators. The transition states are shown in the inset, with the marks of linear and curved arrows to indicate the migration pathway for the H atom.}
    \label{fig7}
\end{figure}

On the other hand, the inherent noise in GPR models can sometimes be beneficial, particularly in preventing convergence to shallow local minima during NEB pathway optimization. In practical NEB calculations, the optimized trajectory heavily depends on the initial guess. For example, in the second step of H$_2$S dissociation (HS + H$\rightarrow$ 2H + S), a linear dissociation path is often assumed as the initial guess. Figure \ref{fig7} shows two distinct pathways found by DFT and GPR calculations. Starting from a linear initial guess, the DFT-driven NEB converges to a linear pathway with a 0.46 eV activation barrier after 53 iterations. In contrast, our GPR model's uncertainty-driven exploration, also explored the linear pathway in the beginning, but then revealed a curved pathway with a lower activation barrier of 0.31 eV after 160 iterations. This serendipitous finding suggests that the stochastic nature of GPR predictions can occasionally help discover unexpected yet physically meaningful reaction pathways. 

\section{Conclusion}
\label{conclusion}
In this work, we introduce \texttt{GPR\_calculator}, a Python and C++ package designed to construct on-the-fly surrogate models using Gaussian Process Regression to expedite computationally intensive electronic structure calculations. A key feature of \texttt{GPR\_calculator} is its ability to dynamically train a GPR model during simulations, accurately predicting energies and forces while quantifying uncertainty. The package is designed for ease of use and seamless integration with the \texttt{ASE}. We have demonstrated the capabilities of \texttt{GPR\_calculator} through benchmark examples, showcasing its acceleration of NEB calculations for surface diffusion and reactions. Our results indicate that \texttt{GPR\_calculator} can substantially decrease the computational demands of these simulations without sacrificing accuracy. 

The framework is designed to be extensible, allowing users to easily add new calculators other than VASP (e.g., Quantum Espresso \cite{QE-2009}, DFTB+ \cite{dftb+}), new NEB optimization algorithms (e.g., L-BFGS), as well as new GPR functionalities (e.g, new descriptors \cite{bartok2013on}, new kernel choices \cite{bishop2006pattern}). In addition, the training data collected from the on-the-fly simulation can also be used for developing other flavors of machine learning force fields such as Neural Networks \cite{NEB-ML-2016}. In the future, we will focus on improving the efficiency of the GPR model, particularly in terms of scaling to larger datasets and enhancing the accuracy of predictions. We anticipate that \texttt{GPR\_calculator} will serve as a valuable asset for researchers in computational materials science and catalysis, enabling more efficient exploration of complex chemical processes.

\section*{Acknowledgments}
This research was sponsored by the U.S. Department of Energy, Office of Science, Office of Basic Energy Sciences and the Established Program to Stimulate Competitive Research (EPSCoR) under the DOE Early Career Award No. DE-SC0024866. The computing resources are provided by ACCESS (TG-DMR180040). The authors thank Dr. Artur Wolek (UNC Charlotte) and Dr. Yu Lin (SLAC) for insightful discussions. During the preparation of this work the authors used GitHub Copilot in order to improve the code readability and documentation. After using this tool/service, the authors reviewed and edited the content as needed and take full responsibility for the content of the publication.

\section*{}

\bibliographystyle{elsarticle-num}
\bibliography{2025-cpc-gpr}
\end{document}